\documentclass[12pt]{article}
\usepackage{graphicx}
\usepackage{amsmath}
\usepackage{amssymb}
\usepackage{caption2}
\begin{document}
\bibliographystyle{prsty}
\begin{center}
{\large {\bf \sc{  Scalar form-factors
$f_{\pi\pi}(Q^2)$  and $f_{KK}(Q^2)$ with light-cone QCD sum rules  }}} \\[2mm]
Zhi-Gang Wang \footnote{E-mail,wangzgyiti@yahoo.com.cn.  }  \\
 Department of Physics, North China Electric Power University,
Baoding 071003, P. R. China
\end{center}

\begin{abstract}
In this article, we calculate the scalar form-factors
$f_{\pi\pi}(Q^2)$ and $f_{KK}(Q^2)$ in the framework of the
light-cone QCD sum rules approach. The numerical value of the
$f_{\pi\pi}(Q^2)$  changes  quickly with   variation of  $Q^2$ near
zero momentum transfer, while  the $f_{KK}(Q^2)$ has rather good
behavior at small momentum transfer. The   value $f_{KK}(0)
=2.21^{+0.35}_{-0.19} GeV$ is compatible  with the result  from the
leading order chiral perturbation theory.  At large momentum
transfer with $Q^2> 6GeV^2$, the form-factor $f_{\pi\pi}(Q^2)$ takes
up the asymptotic behavior  $\frac{1}{Q^2}$ approximately, while the
$f_{KK}(Q^2)$ decreases  more quickly than $\frac{1}{Q^2}$.
\end{abstract}


\section{Introduction}
In the standard model, the  gauge symmetry $SU(2)\times U(1)$ is
spontaneously broken down by the nonvanishing  vacuum expectation
value $v$ of the Higgs field, the fermions obtain their masses
through  Yukawa couplings with the Higgs field. For the light mass
Higgs, the main decay channels maybe  $\pi\pi$ and $\mu\mu$,
although the Yukawa coupling is very small $\sim \frac{1}{v}$
\cite{Donoghue90}, the scalar form-factor $f_{\pi\pi}(t)$ enters the
process $H \to \pi \pi$. However, the scalar form-factor
$f_{\pi\pi}(t)$ is a highly nonperturbative quantity, not a directly
measurable quantity. Omnes representation and Watson theorem can
relate it with $\pi\pi$ and $KK$ scattering data in the spin $J=0$
and isospin $I=0$ channel \cite{CHPT}. It is not unexpected, in the
timelike region,
\begin{eqnarray}
\langle \pi \pi|\bar{u}u+\bar{d}d |0\rangle&=&\langle \pi
\pi|\bar{u}u+\bar{d}d |0\rangle+\langle \pi \pi|\mathcal {T}|\pi
\pi\rangle\langle \pi \pi|\bar{u}u+\bar{d}d |0\rangle +\nonumber \\
&&\langle \pi \pi|\mathcal {T}|K K\rangle\langle
KK|\bar{u}u+\bar{d}d |0\rangle +\cdots \, .
\end{eqnarray}
The scattering matrix elements $\mathcal {T}$ have copious
information and  can be confronted with the experimental data. The
scalar form-factor $f_{\pi\pi}(t)$ has been calculated with the
chiral perturbation theory up to two-loop order now \cite{CHPT}. In
the limit $t=0$,
$\langle\pi(p)|m_u\bar{u}u+m_d\bar{d}d|\pi(p)\rangle$ ($=m_\pi^2$)
and  $\langle K(p)|m_u\bar{u}u+m_s\bar{s}s|K(p)\rangle$ ($=m_K^2$)
are often referred to  as   $\sigma$ terms of the mesons $\pi$ and
$K$. Just like  $\sigma$ terms of the nucleons \cite{Reya74}, they
can put a severe constraint on the  scalar form-factors at zero
momentum transfer.

Semileptonic decays $K\to\pi\ell \nu$ ($K_{\ell 3}$)  provide the
most precise determination of the Cabibbo-Kobayashi-Maskawa (CKM)
matrix element $|V_{us}|$ \cite{CKM}. The experimental input
parameters are the semileptonic decay widths and the vector
form-factors $f^+_{K\pi}(q^2)$ and $f^-_{K\pi}(q^2)$, which  are
necessary in calculating the phase space integrals. The main
uncertainty in the quantity $|V_{us}f^+_{K\pi}(0)|$ comes from the
unknown shape of the hadronic form-factor $f^+_{K\pi}(q^2)$, which
is measurable at $m_l^2<q^2<(m_K-m_\pi)^2$ in  $K_{\ell 3}$
 decays or at $(m_K+m_\pi)^2<q^2<m^2_\tau$  in  $\tau \to K\pi\nu$
decays. The experimental data can be fitted to the functions with
either pole models or series expansions; though systematic errors
are introduced due to the different parameterizations.

In the limit $t=0$, the scalar form- factor  $f_{K\pi}(0)$ has the
value
$f_{K\pi}(0)=\frac{m_K^2-m_\pi^2}{m_s-m_u}f^0_{K\pi}(0)=\frac{m_K^2-m_\pi^2}{m_s-m_u}f^+_{K\pi}(0)$.
Conservation of the vector current implies $f^+_{K\pi}(0)=1$ at zero
momentum transfer \cite{Leutwyler84}.  If the $SU(3)$ symmetry
breaking effects in the scalar channels are  small, the scalar
form-factors $f_{\pi\pi}(0)$ and $f_{KK}(0)$ would have the value
about $1.7GeV$, which is also expected from the leading order chiral
perturbation theory \cite{CHPT}.

In this article, we calculate the value of the scalar form-factor
 $f_{\pi\pi}(Q^2)$ (and $f_{KK}(Q^2)$ as byproduct) in the framework of the light-cone QCD sum
rules approach. In previous works, the scalar form-factors of the
nucleons, which relate with the $\sigma$ terms of the nucleons, have
been calculated with the light-cone QCD sum rules approach
\cite{Wang06}. The light-cone QCD sum rules approach carries out
operator product expansion near the light-cone, $x^2\approx 0$,
instead of the short distance, $x\approx 0$, while the
nonperturbative matrix elements are parameterized by the light-cone
distribution amplitudes (which are classified according to their
twists)  instead of
 the vacuum condensates \cite{LCSR,LCSRreview}. The nonperturbative
 parameters in the light-cone distribution amplitudes are calculated by
 the conventional QCD  sum rules
 and the  values are universal \cite{SVZ79}.

The article is arranged as: in Section 2, we derive the scalar
form-factors  $f_{\pi\pi}(Q^2)$ and $f_{KK}(Q^2)$   with the
light-cone QCD sum rules approach; in Section 3, the numerical
results and discussions; and in Section 4, conclusions.

\section{Scalar  form-factors  $f_{\pi\pi}(Q^2)$ and $f_{KK}(Q^2)$   with light-cone QCD sum rules}

In the following, we write down the definitions  for the scalar
form-factors  $f_{\pi\pi}(q^2)$ and $f_{KK}(q^2)$:
\begin{eqnarray}
\langle \pi(q+p)|\bar{u}(0) u (0)+\bar{d}(0) d
(0)|\pi(p)\rangle&=&2f_{\pi\pi}(q^2)\, , \nonumber \\
 \langle
K(q+p)|\bar{s}(0) s(0)|K(p)\rangle&=&f_{KK}(q^2) \, .
\end{eqnarray}
 We study the scalar form-factors $f_{\pi\pi}(q^2)$ and $f_{KK}(q^2)$ with the
 two-point correlation functions $\Pi_{\mu}^\pi(p,q)$ and
 $\Pi_{\mu}^K(p,q)$ respectively,
\begin{eqnarray}
\Pi_{\mu}^{\pi}(p,q)&=&i \int d^4x \, e^{-i q \cdot x} \,
\langle 0 |T\left\{J^{\pi}_\mu(0) J_d(x)\right\}|\pi(p)\rangle \, , \\
\Pi_{\mu}^K(p,q)&=&i \int d^4x \, e^{-i q \cdot x} \,
\langle 0 |T\left\{J^K_\mu(0) J_s(x)\right\}|K(p)\rangle \, , \\
J^{\pi}_\mu(x)&=&{\bar u}(x)\gamma_\mu \gamma_5  d(x)\, ,\nonumber \\
J^{K}_\mu(x)&=&{\bar u}(x)\gamma_\mu \gamma_5 s(x)\, ,\nonumber\\
J_{d}(x)&=& \bar{d}(x) d(x) \, , \nonumber\\
 J_{s}(x)&=& \bar{s}(x) s(x) \, ,
\end{eqnarray}
 where the axial-vector currents $J^{\pi}_\mu(x)$ and $J^K_\mu(x)$ interpolate    the $\pi$ and $K$ mesons respectively.
 The correlation functions
$\Pi^{P}_{\mu}(p,q)$ (thereafter the $P$ denotes the pseudoscalar
mesons $K$ and $\pi$) can be decomposed as follows:
\begin{eqnarray}
\Pi^{P}_{\mu}(p,q)&=&i\Pi^{P}_{p}\left(q^2,(q+p)^2\right)p_{\mu}+i\Pi^{P}_{q}
\left(q^2,(q+p)^2\right)q_{\mu} \, ,
\end{eqnarray}
due to  Lorentz covariance.  In this article, we derive  the sum
rules with the tensor structures $p_\mu$ and $q_\mu$ respectively.

According to the basic assumption of current-hadron duality in the
QCD sum rules approach \cite{SVZ79}, we can insert  a complete
series of intermediate states with the same quantum numbers as the
current operators $J^P_\mu(x)$   into the correlation functions
$\Pi^P_{\mu}(p,q) $  to obtain the hadronic representation. After
isolating the ground state contributions from the pole terms of the
$\pi$ and   $K$  mesons, the correlation functions $ \Pi^P_\mu(p,q)$
can be expressed in the following form:
\begin{eqnarray}
\Pi_{\mu}^P(p,q)&=&\frac{if_P
}{m_P^2-(q+p)^2}\left\{f^p_{PP}(q^2)p_\mu+f^q_{PP}(q^2)q_\mu\right\}+\cdots
\, ,
\end{eqnarray}
where we have not shown the contributions from the high resonances
and continuum states explicitly, they are suppressed after Borel
transformation and subtraction. We introduce up-indexes  $p$ and $q$
to denote the form-factors with the tensor structures $p_\mu$ and
$q_\mu$ respectively.
 We use the standard definitions  for the weak  decay constants $f_P$,
\begin{eqnarray}
\langle0|J^{P}_\mu(0)|P(p)\rangle&=& if_{P}p_\mu  \,  .\nonumber
\end{eqnarray}

 In the following, we briefly outline
operator product expansion for the correlation functions
$\Pi^P_\mu(p,q)$
 in perturbative QCD theory. The calculations are
performed at  large spacelike momentum regions $P^2=-(q+p)^2\gg 0$
and  $Q^2=-q^2\gg 0$, which correspond to small light-cone distance
$x^2\approx 0$   required by  validity of the operator product
expansion approach. We write down the propagator of a massive quark
in the external gluon field in Fock-Schwinger gauge first
\cite{Belyaev94}:
\begin{eqnarray}
&&\langle 0 | T \{q_i(x_1)\, \bar{q}_j(x_2)\}| 0 \rangle =
 i \int\frac{d^4k}{(2\pi)^4}e^{-ik(x_1-x_2)}\nonumber\\
 &&\left\{
\frac{\not\!k +m}{k^2-m^2} \delta_{ij} -\int\limits_0^1 dv\, g_s \,
G^{\mu\nu}_{ij}(vx_1+(1-v)x_2)
 \right. \nonumber \\
&&\left. \Big[ \frac12 \frac {\not\!k
+m}{(k^2-m^2)^2}\sigma_{\mu\nu} - \frac1{k^2-m^2}v(x_1-x_2)_\mu
\gamma_\nu \Big]\right\}\, ,
\end{eqnarray}
where  $G_{\mu \nu }$ is the gluonic field strength, $g_s$ denotes
the strong coupling constant. Substituting the above $d$, $s$ quark
propagators and the corresponding $\pi$, $K$ mesons light-cone
distribution amplitudes into the correlation functions $\Pi^\pi_\mu$
 and $\Pi^K_\mu$ in Eqs.(3-4), and completing the integrals over the variables $x$
and $k$, finally we obtain the representations at the level of
quark-gluon degrees of freedom. The explicit expressions are given
in the appendix.

 In calculation, we
have used the two-particle and three-particle $K$ and $\pi$ mesons
light-cone distribution amplitudes \cite{LCSR,LCSRreview,Belyaev94}.
The explicit expressions of the $K$ meson light-cone distribution
amplitudes are
  presented in the appendix,  the corresponding ones for
   the $\pi$ meson can be obtained by simple substitution
  of the nonperturbative parameters. The parameters in the
light-cone distribution amplitudes are scale dependent and can be
estimated with the QCD sum rules approach
\cite{LCSR,LCSRreview,Belyaev94}. In this article, the energy scale
$\mu$ is chosen to be  $\mu=1GeV$.

We take   Borel transformation with respect to  the variable
$P^2=-(q+p)^2$    for the correlation functions $\Pi_p^{P}$ and
$\Pi_q^{P}$, and obtain the analytical expressions for those
invariant functions. After matching  with the hadronic
representations below the thresholds, we obtain the following four
sum rules for the scalar form-factors $f^p_{PP}(q^2)$ and
$f^q_{PP}(q^2)$:

\begin{eqnarray}
f^p_{\pi\pi}(q^2) &=&  \frac{ m_\pi^2}{m_u+m_d}\int_{\Delta_\pi}^1du
\phi_p(u)e^{-DD}\nonumber\\
&&-m_d m_\pi^2\int_{\Delta_\pi}^1du \int_0^u
dt \frac{B(t)}{uM^2}e^{-DD}  \nonumber\\
&&-\frac{1}{6}\frac{ m_\pi^2}{m_u+m_d}\int_{\Delta_\pi}^1du
\phi_\sigma(u)\left\{\left[1-u\frac{d}{du}
\right]\frac{1}{u}+\frac{2m_d^2}{u^2M^2}\right\}e^{-DD}
  \nonumber\\
  &&+m_d \int_{\Delta_\pi}^1 du \left\{ \frac{\phi_\pi(u)}{u} - \frac{m_\pi^2m_d^2A(u)}{4u^3M^4}\right\}e^{-DD}\nonumber\\
&&+\frac{f_{3\pi}}{f_{\pi}}\int_0^1dv \int_0^1d\alpha_g
\int_0^{1-\alpha_g}d\alpha_d
\varphi_{3\pi}(\alpha_u,\alpha_g,\alpha_d)\Theta(u-\Delta_\pi)\nonumber\\
&&\left\{\frac{(2v-3) m_\pi^2
}{uM^2}+2v\frac{d}{du}\frac{1}{u}\right\}e^{-DD}\mid_{u=(1-v)\alpha_g+\alpha_d}
\nonumber\\
&&-2m_d  m_\pi^4\int_0^1dv v \int_0^1 d\alpha_g\int_0^{\alpha_g}
d\beta\int_0^{1-\beta}d\alpha \nonumber\\
&&\frac{\Phi(1-\alpha-\beta,\beta,\alpha)\Theta(u-\Delta_\pi)}{u^2M^4}e^{-DD}\mid_{u=1-v\alpha_g}
\nonumber \\
&& +2m_d  m_\pi^4\int_0^1 dv\int_0^1 d\alpha_g\int_0^{1-\alpha_g}
d\alpha_d
 \int_0^{\alpha_d}d\alpha \nonumber\\
 &&\frac{\Phi(1-\alpha-\alpha_g,\alpha_g,\alpha)\Theta(u-\Delta_\pi)}
{u^2M^4}e^{-DD}\mid_{u=(1-v)\alpha_g+\alpha_d}\nonumber\\
&&+m_d   m_\pi^2  \int_0^1dv \int_0^1d\alpha_g
\int_0^{1-\alpha_g}d\alpha_d \nonumber\\
&&\frac{\Psi(\alpha_u,\alpha_g,\alpha_d)\Theta(u-\Delta_\pi)}{u^2M^2}e^{-DD}\mid_{u=(1-v)\alpha_g+\alpha_d}
\, ,
\end{eqnarray}

\begin{eqnarray}
 f^q_{\pi\pi}(q^2)  &=&  \frac{
m_\pi^2}{m_u+m_d}\int_{\Delta_\pi}^1du
\frac{\phi_p(u)}{u}e^{-DD}\nonumber\\
&&-m_d  m_\pi^2\int_{\Delta_\pi}^1du \int_0^u
dt\frac{B(t)}{u^2M^2}e^{-DD} \nonumber\\
&&+\frac{1}{6}\frac{  m_\pi^2}{m_u+m_d}\int_{\Delta_\pi}^1du
\phi_\sigma(u)\frac{d}{du} \frac{1}{u}e^{-DD}
  \nonumber\\
  &&+\frac{f_{3\pi}}{f_\pi}m_\pi^2\int_0^1dv \int_0^1d\alpha_g
  \int_0^{1-\alpha_g}d\alpha_d
\varphi_{\pi}(\alpha_u,\alpha_g,\alpha_d)\nonumber\\
&&\Theta(u-\Delta_\pi)\frac{2v-3
}{u^2M^2}e^{-DD}\mid_{u=(1-v)\alpha_g+\alpha_d}
\nonumber\\
&&-2m_d  m_\pi^4\int_0^1dv v \int_0^1 d\alpha_g\int_0^{\alpha_g}
d\beta\int_0^{1-\beta}d\alpha \nonumber \\
&& \frac{\Phi(1-\alpha-\beta,\beta,\alpha)
\Theta(u-\Delta_\pi)}{u^3M^4}e^{-DD}\mid_{1-v\alpha_g}
\nonumber \\
&& +2m_d  m_\pi^4\int_0^1 dv\int_0^1 d\alpha_g\int_0^{1-\alpha_g}
d\alpha_d
 \int_0^{\alpha_d}d\alpha \nonumber\\
 &&\frac{\Phi(1-\alpha-\alpha_g,\alpha_g,\alpha)\Theta(u-\Delta_\pi)}{u^3M^4}e^{-DD}\mid_{u=(1-v)\alpha_g+\alpha_d}
\, ,
\end{eqnarray}

\begin{eqnarray}
  f^p_{KK}(q^2)
 &=&  \frac{ m_K^2}{m_u+m_s}\int_{\Delta_K}^1du
\phi_p(u)e^{-EE}\nonumber\\
&&-m_s m_K^2\int_{\Delta_K}^1du \int_0^u
dt \frac{B(t)}{uM^2}e^{-EE}  \nonumber\\
&&-\frac{1}{6}\frac{ m_K^2}{m_u+m_s}\int_{\Delta_K}^1du
\phi_\sigma(u)\left\{\left[1-u\frac{d}{du}
\right]\frac{1}{u}+\frac{2m_s^2}{u^2M^2}\right\}e^{-EE}
  \nonumber\\
  &&+m_s\int_{\Delta_K}^1 du \left\{ \frac{\phi_K(u)}{u} - \frac{m_K^2m_s^2A(u)}{4u^3M^4}\right\}e^{-EE}\nonumber\\
&&+\frac{f_{3K}}{f_K}\int_0^1dv \int_0^1d\alpha_g
\int_0^{1-\alpha_g}d\alpha_s
\varphi_{3K}(\alpha_u,\alpha_g,\alpha_s)\Theta(u-\Delta_K)\nonumber\\
&&\left\{\frac{(2v-3) m_K^2
}{uM^2}+2v\frac{d}{du}\frac{1}{u}\right\}e^{-EE}\mid_{u=(1-v)\alpha_g+\alpha_s}
\nonumber\\
&&-2m_s  m_K^4\int_0^1dv v \int_0^1 d\alpha_g\int_0^{\alpha_g}
d\beta\int_0^{1-\beta}d\alpha \nonumber\\
&&\frac{\Phi(1-\alpha-\beta,\beta,\alpha)\Theta(u-\Delta_K)}{u^2M^4}e^{-EE}\mid_{u=1-v\alpha_g}
\nonumber \\
&& +2m_s  m_K^4\int_0^1 dv\int_0^1 d\alpha_g\int_0^{1-\alpha_g}
d\alpha_s
 \int_0^{\alpha_s}d\alpha \nonumber\\
 &&\frac{\Phi(1-\alpha-\alpha_g,\alpha_g,\alpha)\Theta(u-\Delta_K)}
{u^2M^4}e^{-EE}\mid_{u=(1-v)\alpha_g+\alpha_s}\nonumber\\
&&+m_s   m_K^2  \int_0^1dv \int_0^1d\alpha_g
\int_0^{1-\alpha_g}d\alpha_s \nonumber\\
&&\frac{\Psi(\alpha_u,\alpha_g,\alpha_s)\Theta(u-\Delta_K)}{u^2M^2}e^{-EE}\mid_{u=(1-v)\alpha_g+\alpha_s}
\, ,
\end{eqnarray}

\begin{eqnarray}
f^q_{KK}(q^2) &=&  \frac{ m_K^2}{m_u+m_s}\int_{\Delta_K}^1du
\frac{\phi_p(u)}{u}e^{-EE}\nonumber \\
&&-m_s m_K^2\int_{\Delta_K}^1du \int_0^u
dt\frac{B(t)}{u^2M^2}e^{-EE} \nonumber\\
&&+\frac{1}{6}\frac{ m_K^2}{m_u+m_s}\int_{\Delta_K}^1du
\phi_\sigma(u)\frac{d}{du} \frac{1}{u}e^{-EE}
  \nonumber\\
  &&+\frac{f_{3K}}{f_K}m_K^2\int_0^1dv \int_0^1d\alpha_g
  \int_0^{1-\alpha_g}d\alpha_s
\varphi_{3K}(\alpha_u,\alpha_g,\alpha_s)\nonumber\\
&&\Theta(u-\Delta_K)\frac{2v-3
}{u^2M^2}e^{-EE}\mid_{u=(1-v)\alpha_g+\alpha_s}
\nonumber\\
&&-2m_s m_K^4\int_0^1dv v \int_0^1 d\alpha_g\int_0^{\alpha_g}
d\beta\int_0^{1-\beta}d\alpha\nonumber\\
&&\frac{\Phi(1-\alpha-\beta,\beta,\alpha)\Theta(u-\Delta_K)}{u^3M^4}e^{-EE}\mid_{1-v\alpha_g}
\nonumber \\
&& +2m_s m_K^4\int_0^1 dv\int_0^1 d\alpha_g\int_0^{1-\alpha_g}
d\alpha_s
 \int_0^{\alpha_s}d\alpha \nonumber\\
 &&\frac{\Phi(1-\alpha-\alpha_g,\alpha_g,\alpha)\Theta(u-\Delta_K)}{u^3M^4}e^{-EE}\mid_{u=(1-v)\alpha_g+\alpha_s}
\, ,
\end{eqnarray}
where
\begin{eqnarray}
DD &=& \frac{m_d^2-u^2m_\pi^2-(1-u)q^2}{uM^2} \, , \nonumber\\
EE &=& \frac{m_s^2-u^2m_K^2-(1-u)q^2}{uM^2} \, , \nonumber\\
\Delta_\pi&=&\frac{m_d^2-q^2}{s_\pi^0-q^2} \, , \nonumber\\
\Delta_K&=&\frac{m_s^2-q^2}{s_K^0-q^2}  \, .
\end{eqnarray}
 Where  the  $s^0_\pi$  and $s^0_K$ are threshold parameters for the interpolating currents
$J^{\pi}_{\mu}(x)$ and $J^{K}_\mu(x)$,   respectively.

\section{Numerical results and discussions}
The input parameters of the light-cone distribution amplitudes are
taken as $\lambda_3=1.6\pm0.4$, $f_{3K}=(0.45\pm0.15)\times
10^{-2}GeV^2$, $\omega_3=-1.2\pm0.7$, $\omega_4=0.2\pm0.1$,
$a_2=0.25\pm 0.15$, $a_1=0.06\pm 0.03$, $\eta_4=0.6\pm0.2$  for the
$K$ meson;   $\lambda_3=0.0$, $f_{3\pi}=(0.45\pm0.15)\times
10^{-2}GeV^2$, $\omega_3=-1.5\pm0.7$, $\omega_4=0.2\pm0.1$,
$a_2=0.25\pm 0.15$, $a_1=0.0 $, $\eta_4=10.0\pm3.0$   for the $\pi$
meson \cite{LCSR,LCSRreview,Belyaev94}; and $m_s=(137\pm 27 )MeV$,
$m_u=m_d=(5.6\pm 1.6)  MeV$, $f_K=0.16GeV$, $f_\pi=0.13GeV$,
$m_K=498MeV$, $m_{\pi} =138MeV$. The threshold parameters are chosen
to be $s^0_{K}=1.1GeV^2$ and $s^0_{\pi}=0.8GeV^2$, which can
reproduce the values of the decay constants $f_K=0.16GeV$ and
$f_{\pi}=0.13GeV$ in the QCD sum rules approach.

The Borel parameters in the four sum rules are taken as
$M^2=(0.6-2.0) GeV^2$, in this region, the values of the
form-factors $f_{\pi\pi}(Q^2)$   and   $f_{KK}(Q^2)$  are rather
stable. In this article, we take the special values $M^2=1.2GeV^2$
for the $f_{\pi\pi}(Q^2)$ and $M^2=1.5GeV^2$ for the $f_{KK}(Q^2)$
in numerical calculations. Although such a definite Borel parameter
cannot take into account some uncertainties,   the predictive power
cannot be impaired qualitatively.

From the four sum rules, we observe that the main contributions come
from the two-particle twist-3 light-cone distribution amplitudes,
not  the twist-2 light-cone distribution amplitudes. The
contributions from the twist-2 light-cone distribution amplitudes
are suppressed by the extra factors of small masses $m_d$ or $m_s$.
For the heavy-light form-factors $B \to \pi, K$, the contributions
from the twist-2 light-cone distribution amplitudes are enhanced by
the extra factor of large mass $m_b$, we can take the chiral limit
for the masses of the mesons $K$ and $\pi$ \cite{BKaonPion}; the
contributions from the two-particle twist-3 light-cone distribution
amplitudes are very small and can   be neglected safely.

The uncertainties of the seven parameters $f_{3K}$($f_{3\pi}$),
$a_2$, $a_1$, $\lambda_3$, $\omega_3$, $\omega_4$ and $\eta_4$  can
only result in small uncertainties for the numerical values. The
main uncertainties come from the two parameters  $m_s$ and
$m_q(m_u,m_d)$; the variations of those parameters can lead to large
changes for the numerical values, we should refine the input
parameters $m_q$ (for $f_{\pi\pi}(Q^2)$) and $m_s$ (for
$f_{KK}(Q^2)$), especially the $m_q$ to improve the predictive
ability. It is a difficult work.

Taking into account all the uncertainties, finally we obtain the
numerical values of the scalar form-factors $f_{\pi\pi}(Q^2)$ and
$f_{KK}(Q^2)$, which  are shown in Figs.1-3.  At zero momentum
transfer,
\begin{eqnarray}
 f^p_{\pi\pi}(0) &=&-0.56^{+0.21}_{-0.21} GeV   \, , \nonumber \\
 f^q_{\pi\pi}(0) &=&0.81^{+0.81}_{-0.52} GeV  \, ,  \nonumber \\
  f^p_{KK}(0) &=&0.32^{+0.24}_{-0.24} GeV  \, ,  \nonumber \\
f^q_{KK}(0) &=&2.21^{+0.35}_{-0.19} GeV \, .
\end{eqnarray}

\begin{figure}
\centering
  \includegraphics[totalheight=6cm,width=7cm]{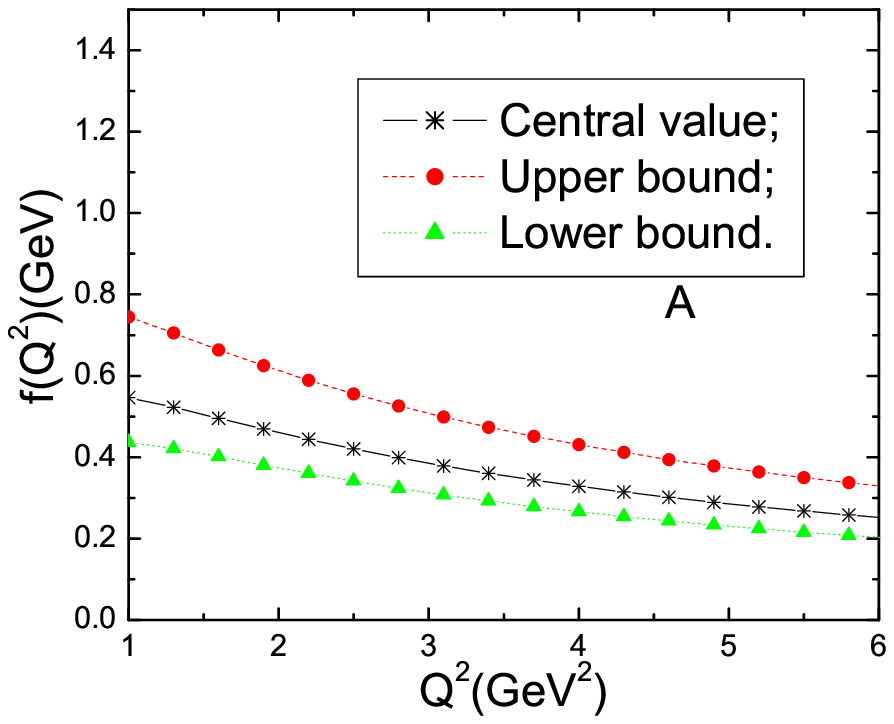}
  \includegraphics[totalheight=6cm,width=7cm]{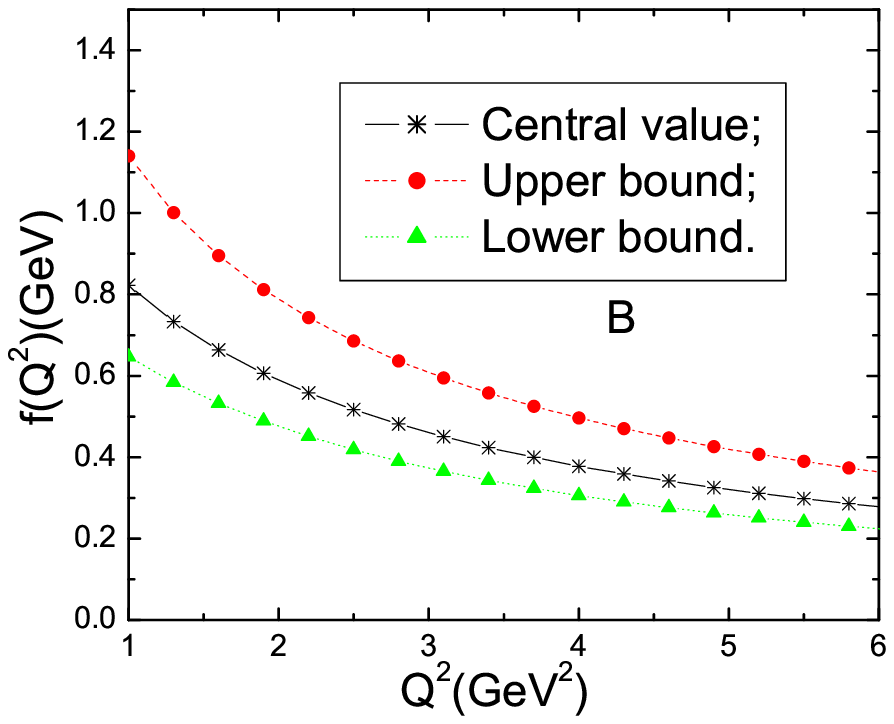}
  \includegraphics[totalheight=6cm,width=7cm]{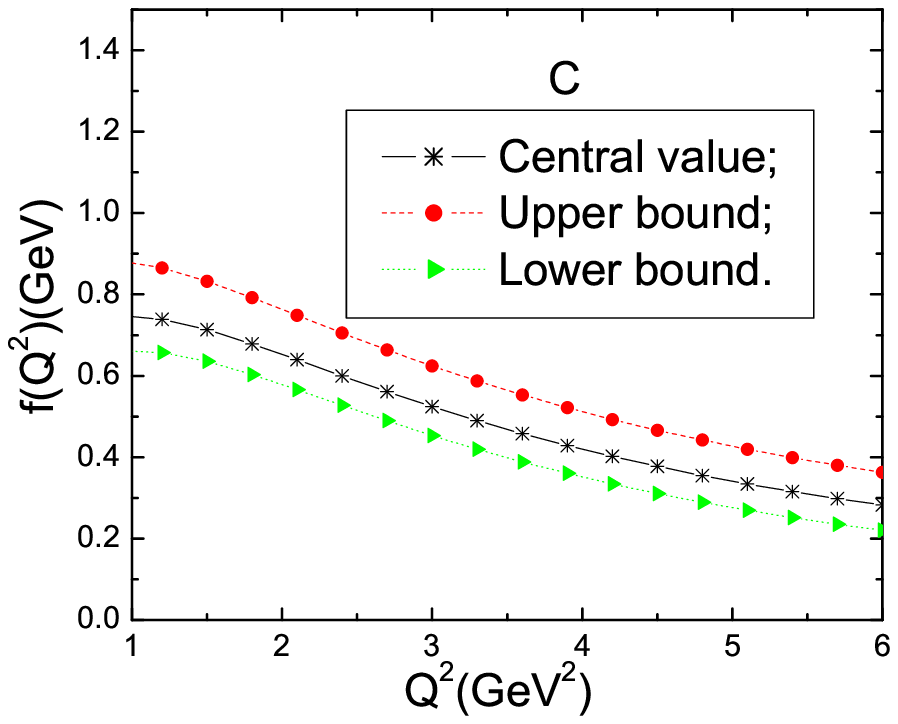}
  \includegraphics[totalheight=6cm,width=7cm]{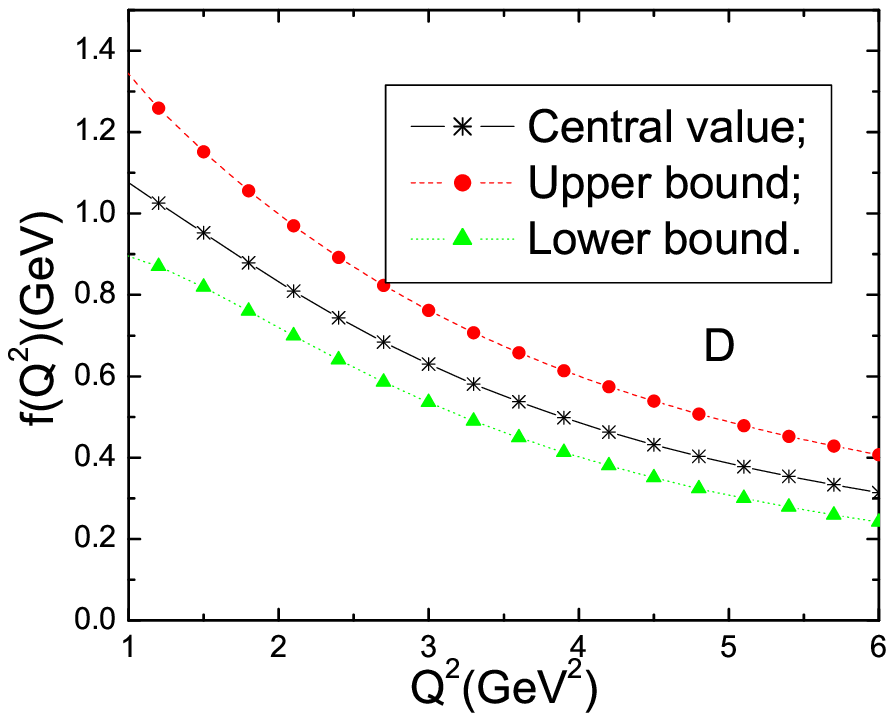}
       \caption{$f^p_{\pi\pi}(Q^2)$(A), $f^q_{\pi\pi}(Q^2)$(B), $f^p_{KK}(Q^2)$(C) and $f^q_{KK}(Q^2)$(D) at the
       range   $Q^2=(1-6)GeV^2$. }
\end{figure}

\begin{figure}
\centering
  \includegraphics[totalheight=6cm,width=7cm]{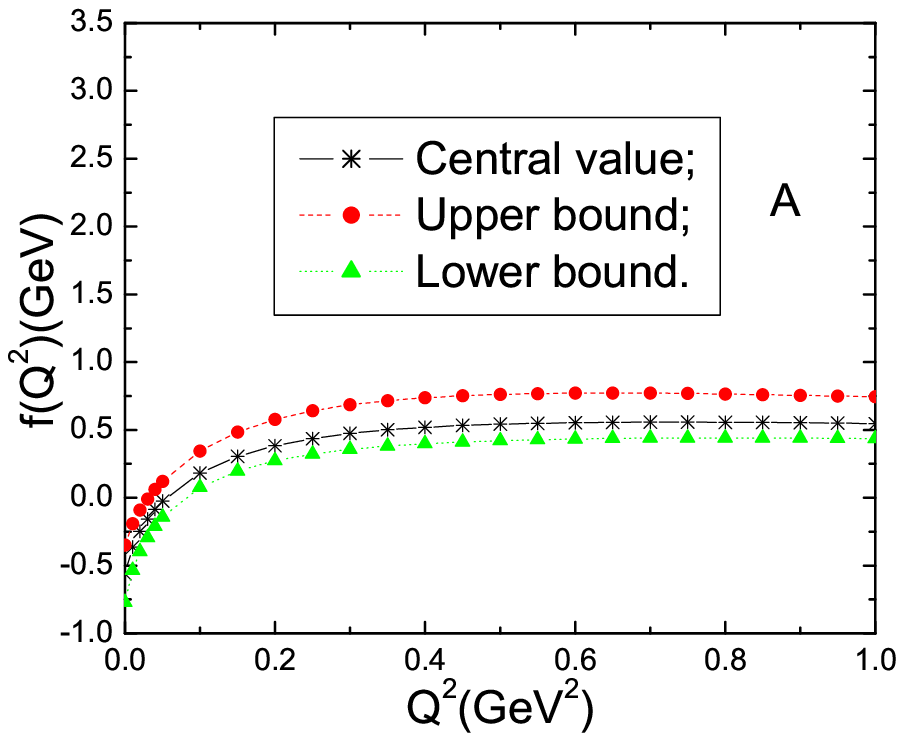}
  \includegraphics[totalheight=6cm,width=7cm]{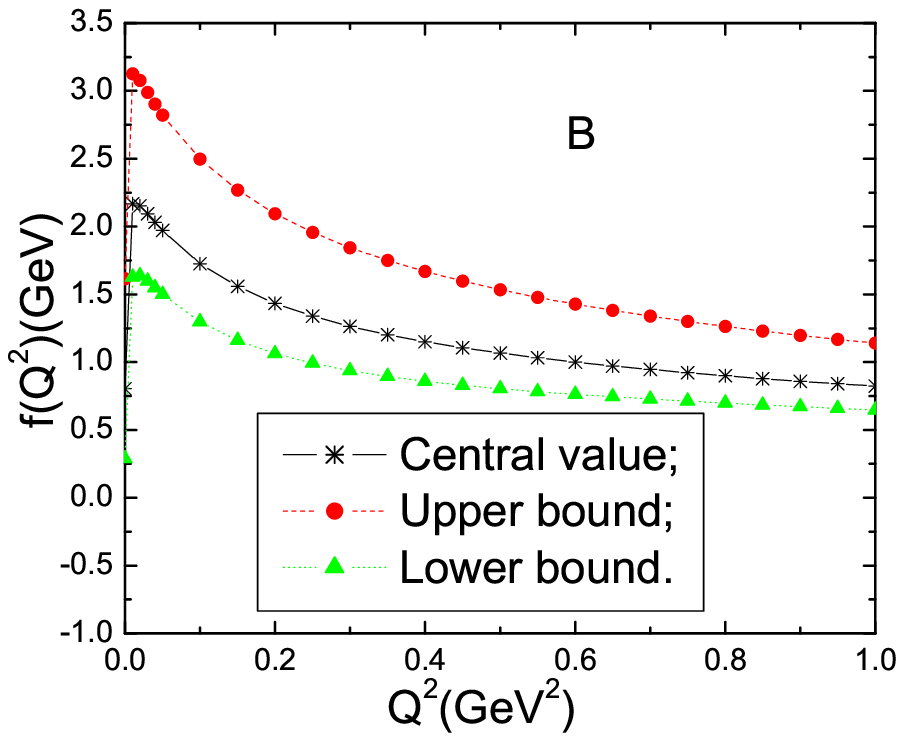}
  \includegraphics[totalheight=6cm,width=7cm]{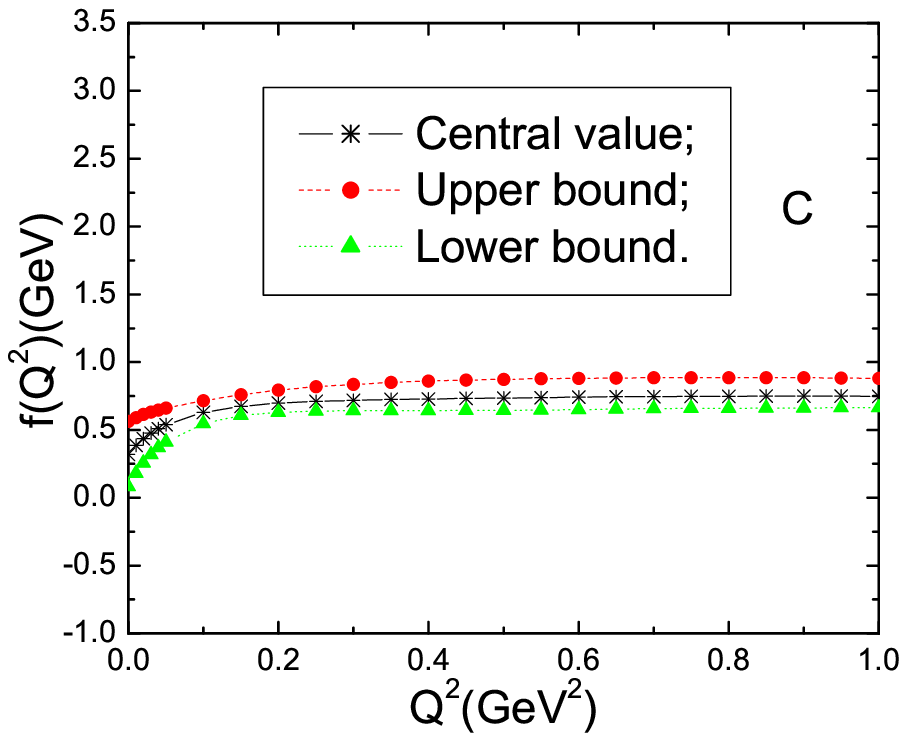}
  \includegraphics[totalheight=6cm,width=7cm]{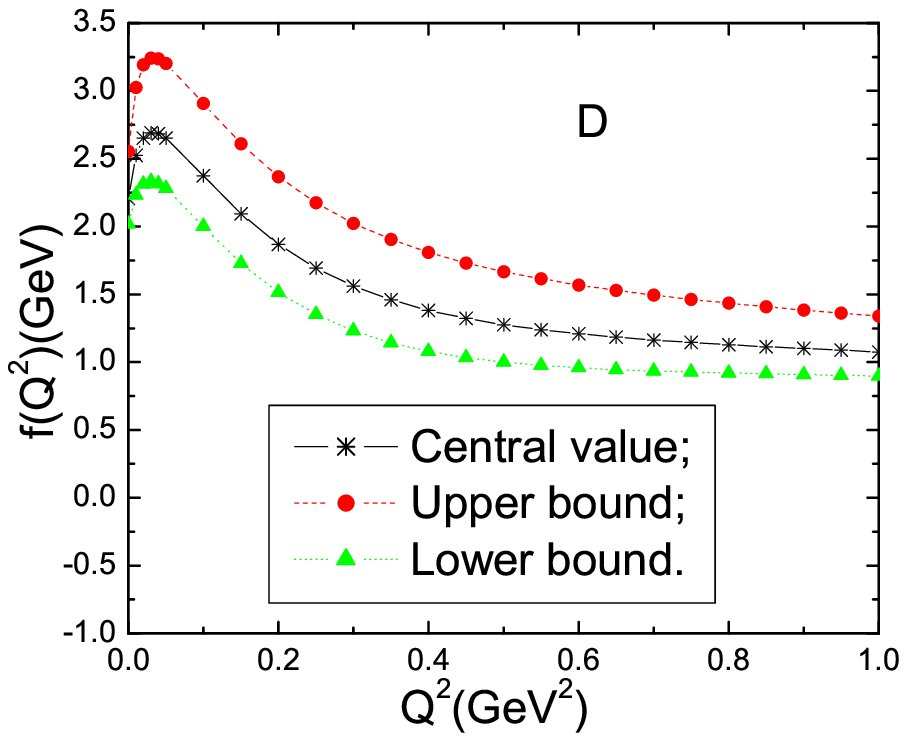}
       \caption{  $f^p_{\pi\pi}(Q^2)$(A), $f^q_{\pi\pi}(Q^2)$(B), $f^p_{KK}(Q^2)$(C) and
       $f^q_{KK}(Q^2)$(D) at the range $Q^2=(0-1)GeV^2$. }
\end{figure}

\begin{figure}
\centering
  \includegraphics[totalheight=8cm,width=10cm]{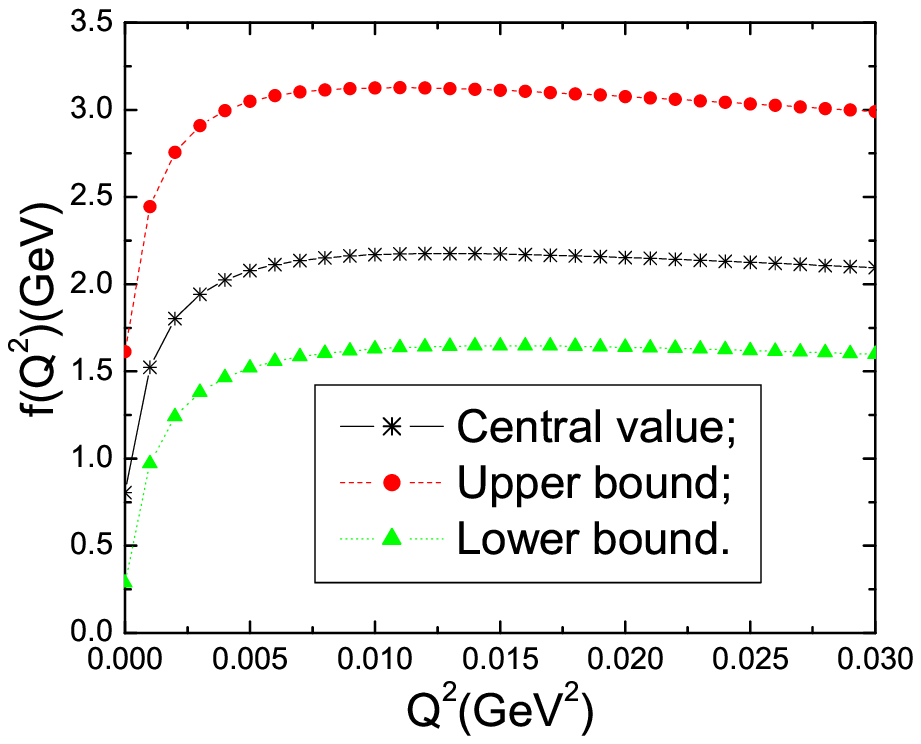}
         \caption{   $f^q_{\pi\pi}(Q^2)$ at the range $Q^2=(0-0.03)GeV^2$. }
\end{figure}

In the light-cone QCD sum rules approach, we carry out operator
product expansion near the light-cone $x^2\approx 0$, which
corresponds to  $Q^2\gg 0$ and $P^2 \gg 0 $. The four sum rules
$f^p_{PP}(Q^2)$ and $f^q_{PP}(Q^2)$ can be taken as some functions
that model the scalar form-factors $ f_{PP}(Q^2) $   at large
momentum transfer, we extrapolate  them to zero momentum transfer
with an analytical continuation.

We can borrow some ideas from the electromagnetic $\pi$-photon
form-factor $f_{\gamma^* \pi^0 }(Q^2)$. The value of
 $f_{\gamma^* \pi^0 }(0)$ is fixed by partial conservation of
 the axial current and the effective  anomaly lagrangian,
$f_{\gamma^* \pi^0 }(0) = \frac1{\pi f_{\pi}}$. In the limit of
large-$Q^2$,   perturbative QCD predicts that $f_{\gamma^* \pi^0
}(Q^2) =\frac{4\pi f_{\pi}}{Q^2} $.
 The Brodsky-Lepage interpolation formula \cite{BJ81}
 \begin{eqnarray}
f_{\gamma^* \pi^0 }(Q^2) = \frac{1}{ \pi f_{\pi} \left [1+Q^2/(4
\pi^2 f_{\pi}^2) \right ]} =\frac{1}{ \pi f_\pi (1+Q^2/s_0) }
\nonumber
\end{eqnarray}
 can reproduce both the value at $Q^2 =0$
  and the  behavior at large-$Q^2$. The energy scale $s_0$ ($s_0 = 4 \pi^2
f_{\pi}^2 \approx 0.67 \, GeV^2$) is numerically   close to   the
squared mass of the $\rho$ meson, $m_{\rho}^2 \approx 0.6 \, GeV^2$.
The Brodsky-Lepage interpolation formula is similar to the result of
the vector meson dominance approach, $f_{\gamma^* \pi^0 }(Q^2) =
1/\left\{\pi f_\pi (1+Q^2/m_{\rho}^2)\right\}$. In the latter case,
the calculation is performed at the timelike energy scale
$q^2<1GeV^2$ and the electromagnetic  current is saturated by the
vector meson $\rho$, where the mass $m_{\rho}$ serves as a parameter
determining the pion charge radius. With a slight modification of
the mass parameter, $m_\rho=\Lambda_\pi=776MeV$, the experimental
data can be well described by the single-pole formula  at the
interval $Q^2=(0-10)GeV^2$ \cite{CLEO97}.

In Ref.\cite{WangSN07}, the four form-factors of  $\Sigma \to n$
have satisfactory behaviors at large $Q^2$, which are expected by
naive power counting rules, and they have finite values at $Q^2=0$.
The analytical expressions of the four form-factors $f_1(Q^2)$,
$f_2(Q^2)$, $g_1(Q^2)$ and $g_2(Q^2)$ are taken as  Brodsky-Lepage
type of interpolation formulae, although they are calculated at
rather large $Q^2$, the extrapolation to lower energy transfer has
no solid theoretical foundation. The numerical values of $f_1(0)$,
$f_2(0)$, $g_1(0)$ and $g_2(0)$ are
 compatible with the  experimental data and  theoretical
 calculations (in magnitude).

 In Ref.\cite{Wang07E}, the vector form-factors $f_{K\pi}^+(Q^2)$
 and  $f_{K\pi}^-(Q^2)$ are
 also taken as Brodsky-Lepage  type of interpolation formulae, the behaviors of low momentum
 transfer  are rather good in some channels.

 In this article, we take the scalar form-factors
 $f^p_{PP}(Q^2)$ and $f^q_{PP}(Q^2)$
   as  Brodsky-Lepage  type of interpolation formulae, unfortunate, the low energy
   behaviors  (for $Q^2<0.03GeV^2$) of  the $f^p_{\pi\pi}(Q^2)$ and  $f^q_{\pi\pi}(Q^2)$
  are rather bad.

  It is obvious that the model functions $f^p_{PP}(Q^2)$ and $f^q_{PP}(Q^2)$ may
  have good or bad low-$Q^2$ behaviors, although
  they have solid theoretical foundation at large momentum
transfer. We extrapolate the model functions tentatively to zero
momentum transfer, systematic errors maybe very large and the
results maybe unreliable. The predictions merely  indicate the
possible values of the light-cone QCD sum rules approach, they
should be confronted with the experimental data or other theoretical
approaches.

In the limit $Q^2=0$, $\Delta_K\approx 0.017$ and
$\Delta_\pi\approx0.00004$. Although the terms proportional to
$\frac{1}{u^n}\exp\{ -DD\}$ and $\frac{1}{u^n}\exp\{ -EE\}$ have
finite values, the contributions from the end-point are greatly
enhanced. Comparing with the $f^p_{PP}(Q^2)$, the $f^q_{PP}(Q^2)$
 have more terms with the extra factor of
$\frac{1}{u}$. It is not unexpected that the $f^q_{PP}(Q^2)$ have
larger values than the corresponding $f^p_{PP}(Q^2)$  at small
momentum transfer, which are shown in Fig.2.

If we take the value $Q^2=(0.01-0.02)GeV^2$,
$\Delta_\pi\approx0.012-0.024$ and $\Delta_K\approx0.026-0.035$.
 The $f^q_{\pi\pi}(Q^2)$ has
larger contributions from the end-point of the light-cone
distribution amplitudes than the $f^q_{KK}(Q^2)$. Without nice
cancelation among the end-point dominating terms, such an infrared
behavior can result in that the $f^q_{\pi\pi}(Q^2)$ changes quickly
with variation of
  $Q^2$ at $Q^2<0.03GeV^2$, which is shown explicitly in Fig.3 (also Fig.2).

  One can adjust the input parameters
  to cancel the infrared enhancement, however,
  the input parameters are calculated with the QCD sum rules approach \cite{LCSR,LCSRreview,Belyaev94},
  they are not free parameters.
    We should introduce extra
phenomenological form-factors (for example, the Sudakov factor
\cite{HnLi}) to suppress the contribution from the end-point. It is
somewhat of fine-tuning.

The vector form-factor $f^+_{K\pi}(q^2)$ and scalar form-factor
$f^0_{K\pi}(q^2)$ (with the relation
$f^0_{K\pi}(q^2)=f^+_{K\pi}(q^2)+\frac{q^2}{m_K^2-m_\pi^2}f^-_{K\pi}(q^2)$)
\footnote{Current algebra predicts the value of the scalar
form-factor $f^0_{K\pi}(\Delta)$ be $f^0_{K\pi}(\Delta)= -f_K/f_\pi$
at  Callan-Treiman point $\Delta=m_K^2-m_\pi^2$ \cite{CT66}.} are
measured in  $K_{\ell 3}$ decays with the squared momentum $q^2$
transfer to the leptons, where $q^2>m_l^2$. The curves (or shapes)
of the form-factors are always parameterized by the linear model,
quadratic model and pole models to carry out the integrals in the
phase space. The normalization is always chosen to be
$f^+_{K\pi}(0)$, i.e.
$f^+_{K\pi}(q^2)=f^+_{K\pi}(0)\left\{1+\lambda_1q^2+\lambda_2q^4+\cdots\right\}$,
etc, the parameters $\lambda_1$, $\lambda_2$, $\cdots$ can be fitted
by   $\chi^2$, etc \cite{ExpKP}.

In the limit $q^2=0$, $f^0_{K\pi}(0)=f^+_{K\pi}(0)\approx 1$. The
vector form-factor $f^+_{K\pi}(Q^2)$ has been calculated by the ChPT
\cite{ChPTKP}, lattice QCD \cite{LattKP}, (light-cone) QCD sum rules
\cite{Wang07E,LCQCDSRKP},
  etc.  If the $SU(3)$ symmetry works well in the scalar channels,
  the values of  the $f_{\pi\pi}(0)$ and  $f_{KK}(0)$ would not differ from
  the value of the scalar form-factor $f_{K\pi}(0)$ greatly,
  $f_{K\pi}(0)=\frac{m_K^2-m_\pi^2}{m_s-m_u}f^0_{K\pi}(0)\approx
  1.7GeV$. The leading order chiral
perturbation theory also predicts that
$f_{\pi\pi}(0)=f_{KK}(0)=f_{K\pi}(0)\approx1.7GeV$ \cite{CHPT}. The
numerical value $f^p_{KK}(0) =2.21^{+0.35}_{-0.19} GeV$ makes sense,
not very bad.

\begin{figure}
\centering
  \includegraphics[totalheight=6cm,width=7cm]{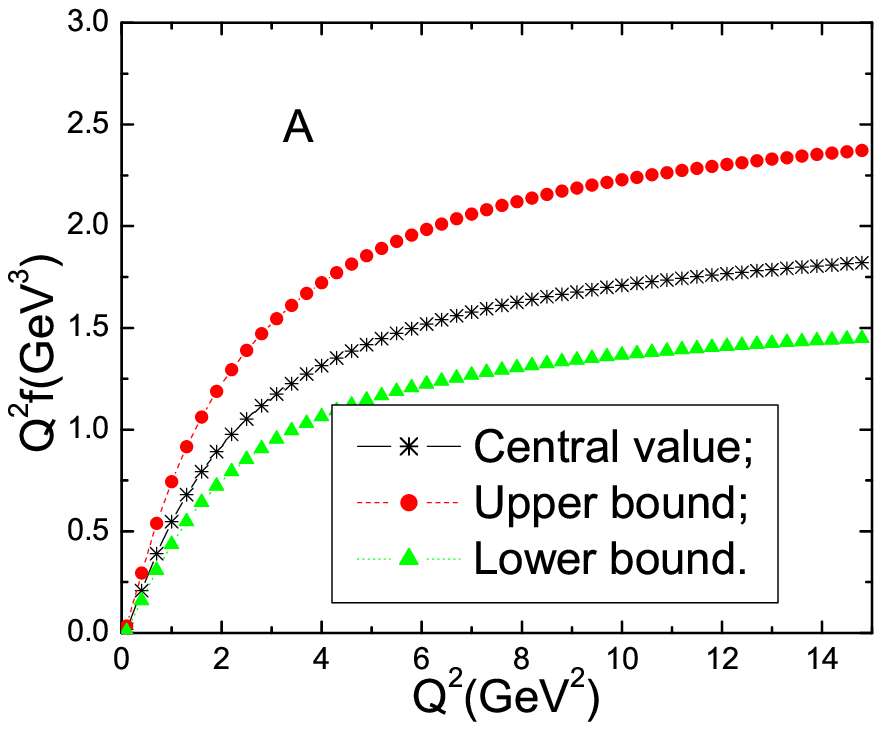}
  \includegraphics[totalheight=6cm,width=7cm]{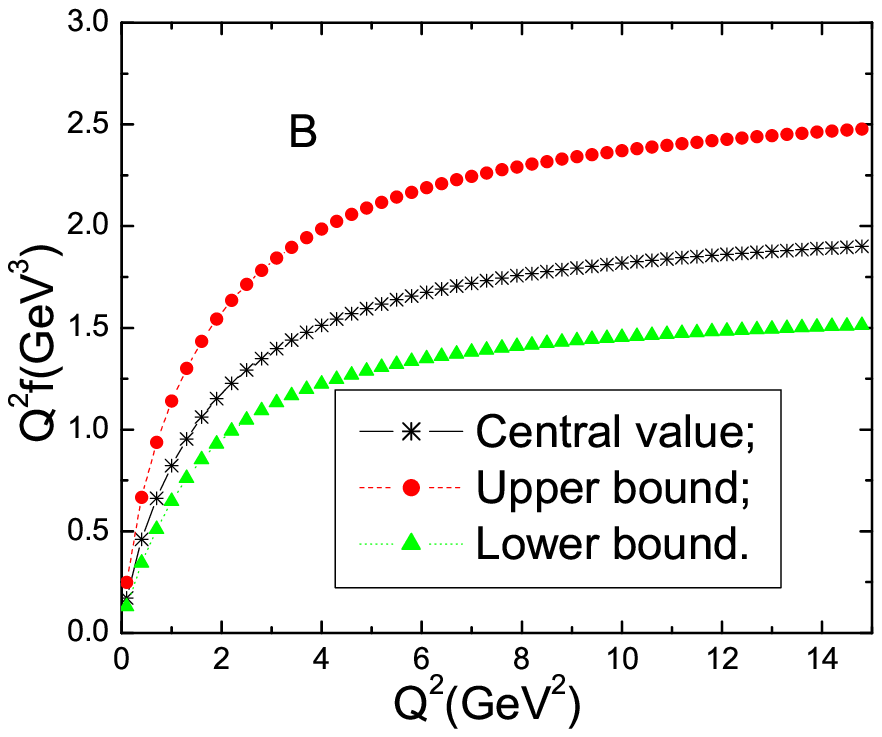}
  \includegraphics[totalheight=6cm,width=7cm]{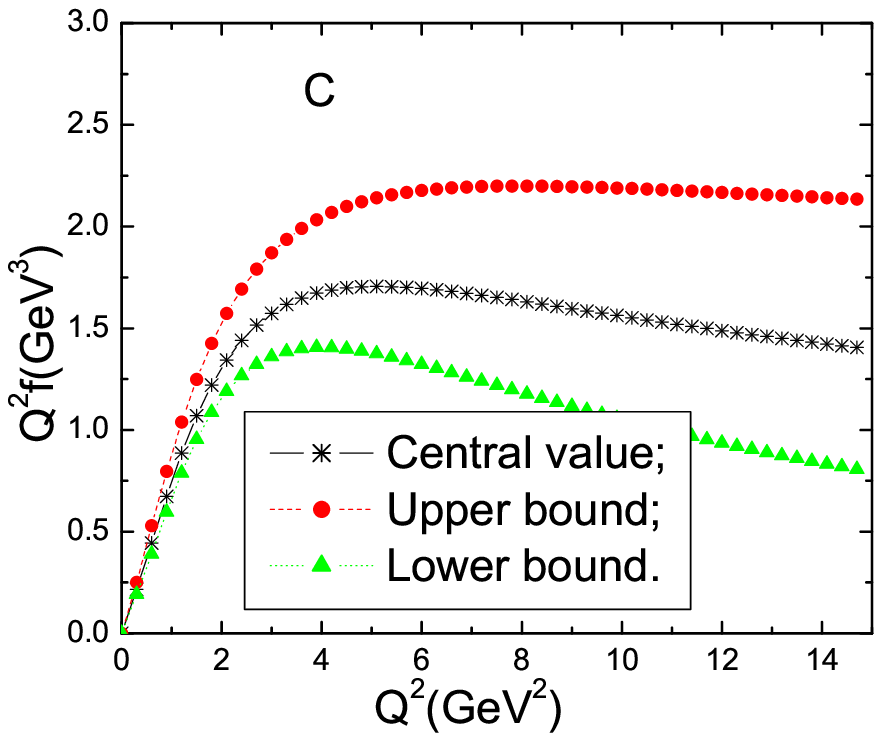}
  \includegraphics[totalheight=6cm,width=7cm]{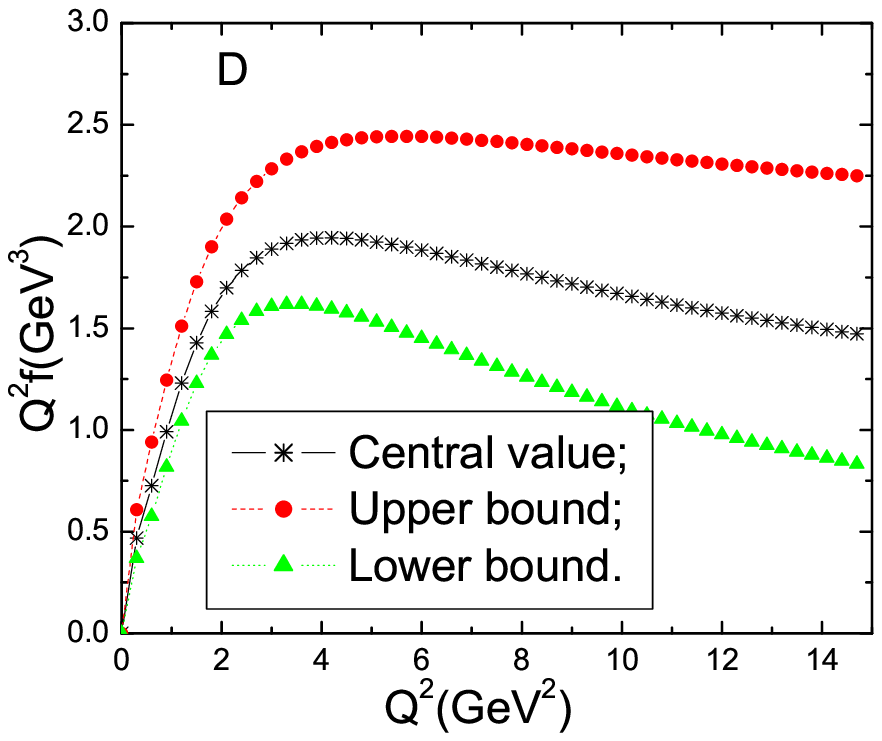}
       \caption{  $Q^2f^p_{\pi\pi}(Q^2)$(A), $Q^2f^q_{\pi\pi}(Q^2)$(B), $Q^2f^p_{KK}(Q^2)$(C) and $Q^2f^q_{KK}(Q^2)$(D)
       at the range $Q^2=(0-15)GeV^2$. }
\end{figure}

In  Fig.4, we plot the form-factors $f^p_{PP}(Q^2)$  and
$f^q_{PP}(Q^2)$ at the range $Q^2=(0-15)GeV^2$. From Fig.4, we can
see that the curves (or shapes) of $Q^2f^p_{\pi\pi}(Q^2)$ and
$Q^2f^q_{\pi\pi}(Q^2)$ are rather flat at $Q^2> 6 GeV^2 $, which
means that at large momentum transfer, the scalar form-factor
$f_{\pi\pi}(Q^2)$ takes up the asymptotic behavior
$f_{\pi\pi}(Q^2)\sim \frac{1}{Q^2}$ approximately. It is expected
from naive power counting rules \cite{Brodsky},  the terms
proportional to $\frac{1}{Q^{2n}}$ with $n\geq 2$ are canceled out
approximately  with each other.
 The scalar form-factor, axial form-factor and induced pseudoscalar
form-factor of the nucleons show the behavior
 $\frac{1}{Q^4}$ at large $Q^2$ \cite{Wang06,WangSN07,WWY}, which is also expected
 from  naive power counting rules \cite{Brodsky}.
 The curves (or shapes) of   $Q^2f^p_{KK}(Q^2)$ and $Q^2f^q_{KK}(Q^2)$  at $Q^2<6GeV^2$ are analogous  to
the electromagnetic form-factors of the $K$ and $\pi$ mesons. At
large momentum transfer with $Q^2>6GeV^2$, the terms of the
$f_{KK}(Q^2)$  proportional to $\frac{1}{Q^{2n}}$ with $n\geq 2$
  manifest themselves, which results in the curves (or
shapes) of  $Q^2f_{KK}(Q^2)$ decreasing  with  increasing  $Q^2$.

The scalar form-factors are complex functions of the input
parameters, in principle, they can be expanded in terms of Taylor
series of $\frac{1}{Q^2}$. At large momentum transfer, for example,
$Q^2=(6-15)GeV^2$,  the central values of the four form-factors can
be fitted numerically as
\begin{eqnarray}
f^p_{\pi\pi}(Q^2)&=& \frac{1.9}{Q^2}-\frac{2.0}{Q^4} \, ,\nonumber \\
f^q_{\pi\pi}(Q^2)&=&\frac{2.0}{Q^2}-\frac{2.1}{Q^4} \, ,\nonumber \\
f^p_{KK}(Q^2)&=& \frac{1.1}{Q^2}+\frac{4.3}{Q^4} \, ,\nonumber \\
f^q_{KK}(Q^2)&=& \frac{1.2}{Q^2}+\frac{4.5}{Q^4} \, .
\end{eqnarray}
The form-factor $f_{\pi\pi}(Q^2)$ has larger $\frac{1}{Q^2}$
dependence and smaller  $\frac{1}{Q^4}$ dependence than the
$f_{KK}(Q^2)$. Although the analytical expressions of
$f^p_{PP}(Q^2)$ (or $f^q_{PP}(Q^2)$) have the same type of $Q^2$
dependence, the coefficients are quantitatively different from each
other due to the $SU(3)$ symmetry breaking effects for the mesons
$\pi$ and $K$. At large momentum transfer,
$\Delta_\pi=\frac{m_d^2+Q^2}{s_\pi^0+Q^2}\approx 1 $ and
$\Delta_K=\frac{m_s^2+Q^2}{s_K^0+Q^2}\approx 1 $, the extra factor
of $\frac{1}{u}$ in the scalar form-factors $f^q_{PP}(Q^2)$  will
not play any significant roles, the form-factor $f^p_{PP}(Q^2)$ and
the corresponding $f^q_{PP}(Q^2)$  approach almost the same form of
$Q^2$ dependence.

\section{Conclusions}

In this article, we calculate the scalar form-factors
$f_{\pi\pi}(Q^2)$ and $f_{KK}(Q^2)$  in the framework of the
light-cone QCD sum rules approach. The scalar form-factor
$f_{\pi\pi}(t)$ enters the light Higgs decay $H \to \pi \pi$, and it
is not a directly  measurable quantity.  The scalar form-factors
$f_{\pi\pi}(0)$ and $f_{KK}(0)$  relate with the $\sigma$ terms of
the $\pi$ and $K$ mesons, respectively. Just like the  $\sigma$
terms of the nucleons, they are highly nonperturbative quantities.
The numerical values of the $f^p_{\pi\pi}(Q^2)$ and
$f^q_{\pi\pi}(Q^2)$ change quickly with the variation of  $Q^2$ near
zero momentum transfer, while the $f^p_{KK}(Q^2)$ and
$f^q_{KK}(Q^2)$ have rather good behaviors  at small momentum
transfer. The value $f^q_{KK}(0)=2.21^{+0.35}_{-0.19}GeV$ is
compatible with the result from the leading order chiral
perturbation theory. At large momentum transfer with $Q^2> 6GeV^2$,
the form-factor $f_{\pi\pi}(Q^2)$ takes up the asymptotic behavior
of $\frac{1}{Q^2}$ approximately, while the $f_{KK}(Q^2)$ decreases
more quickly than $\frac{1}{Q^2}$.

\section*{Appendix}

The light-cone distribution amplitudes of the $K$ meson are defined
as follows:
\begin{eqnarray}
\langle0| {\bar u} (0) \gamma_\mu \gamma_5 s(x) |K(p)\rangle& =& i
f_K p_\mu \int_0^1 du  e^{-i u p\cdot x}
\left\{\phi_K(u)+\frac{m_K^2x^2}{16}
A(u)\right\}\nonumber\\
&&+if_K m_K^2\frac{x_\mu}{2p\cdot x}
\int_0^1 du  e^{-i u p \cdot x} B(u) \, , \nonumber\\
\langle0| {\bar u} (0) i \gamma_5 s(x) |K(p)\rangle &=& \frac{f_K
m_K^2}{ m_s+m_u}
\int_0^1 du  e^{-i u p \cdot x} \phi_p(u)  \, ,  \nonumber\\
\langle0| {\bar u} (0) \sigma_{\mu \nu} \gamma_5 s(x) |K(p)\rangle
&=&i(p_\mu x_\nu-p_\nu x_\mu)  \frac{f_K m_K^2}{6 (m_s+m_u)}
\int_0^1 du
e^{-i u p \cdot x} \phi_\sigma(u) \, ,  \nonumber\\
\langle0| {\bar u} (0) \sigma_{\mu\nu} \gamma_5 g_s
G_{\alpha\beta}(v x)s(x) |K(p)\rangle&=& f_{3 K}\left\{(p_\mu
p_\alpha g^\bot_{\nu
\beta}-p_\nu p_\alpha g^\bot_{\mu \beta}) -(p_\mu p_\beta g^\bot_{\nu \alpha}\right.\nonumber\\
&&\left.-p_\nu p_\beta g^\bot_{\mu \alpha})\right\} \int {\cal
D}\alpha_i \varphi_{3 K} (\alpha_i)
e^{-ip \cdot x(\alpha_s+v \alpha_g)} \, ,\nonumber\\
\langle0| {\bar u} (0) \gamma_{\mu} \gamma_5 g_s G_{\alpha
\beta}(vx)s(x) |K(p)\rangle&=&  p_\mu  \frac{p_\alpha
x_\beta-p_\beta x_\alpha}{p
\cdot x}f_Km_K^2\nonumber\\
&&\int{\cal D}\alpha_i A_{\parallel}(\alpha_i) e^{-ip\cdot
x(\alpha_s +v \alpha_g)}\nonumber \\
&&+ f_Km_K^2 (p_\beta g^\perp_{\alpha\mu}-p_\alpha
g^\perp_{\beta\mu})\nonumber\\
&&\int{\cal D}\alpha_i A_{\perp}(\alpha_i)
e^{-ip\cdot x(\alpha_s +v \alpha_g)} \, ,  \nonumber\\
\langle0| {\bar u} (0) \gamma_{\mu} i g_s \tilde G_{\alpha
\beta}(vx)s(x) |K(p)\rangle&=& p_\mu  \frac{p_\alpha x_\beta-p_\beta
x_\alpha}{p \cdot
x}f_Km_K^2\nonumber\\
&&\int{\cal D}\alpha_i V_{\parallel}(\alpha_i) e^{-ip\cdot
x(\alpha_s +v \alpha_g)}\nonumber \\
&&+ f_Km_K^2 (p_\beta g^\perp_{\alpha\mu}-p_\alpha g^\perp_{\beta\mu})\nonumber\\
&&\int{\cal D}\alpha_i V_{\perp}(\alpha_i) e^{-ip\cdot x(\alpha_s +v
\alpha_g)} \, ,
\end{eqnarray}
where the operator $\tilde G_{\alpha \beta}$  is the dual of
$G_{\alpha \beta}$, $\tilde G_{\alpha \beta}= {1\over 2}
\epsilon_{\alpha \beta  \mu\nu} G^{\mu\nu} $, ${\cal{D}}\alpha_i$ is
defined as ${\cal{D}} \alpha_i =d \alpha_1 d \alpha_2 d \alpha_3
\delta(1-\alpha_1 -\alpha_2 -\alpha_3)$. The  light-cone
distribution amplitudes are parameterized as follows:
\begin{eqnarray}
\phi_K(u,\mu)&=&6u(1-u)
\left\{1+a_1C^{\frac{3}{2}}_1(2u-1)+a_2C^{\frac{3}{2}}_2(2u-1)
+a_4C^{\frac{3}{2}}_4(2u-1)\right\}\, , \nonumber\\
\phi_p(u,\mu)&=&1+\left\{30\eta_3-\frac{5}{2}\rho^2\right\}C_2^{\frac{1}{2}}(2u-1)\nonumber \\
&&+\left\{-3\eta_3\omega_3-\frac{27}{20}\rho^2-\frac{81}{10}\rho^2 a_2\right\}C_4^{\frac{1}{2}}(2u-1)\, ,  \nonumber \\
\phi_\sigma(u,\mu)&=&6u(1-u)\left\{1
+\left[5\eta_3-\frac{1}{2}\eta_3\omega_3-\frac{7}{20}\rho^2-\frac{3}{5}\rho^2 a_2\right]C_2^{\frac{3}{2}}(2u-1)\right\}\, , \nonumber \\
\varphi_{3K}(\alpha_i,\mu) &=& 360 \alpha_u \alpha_s \alpha_g^2
\left \{1 +\lambda_3(\alpha_u-\alpha_s)+ \omega_3 \frac{1}{2} ( 7
\alpha_g
- 3) \right\} \, , \nonumber\\
V_{\parallel}(\alpha_i,\mu) &=& 120\alpha_u \alpha_s \alpha_g \left(
v_{00}+v_{10}(3\alpha_g-1)\right)\, ,
\nonumber \\
A_{\parallel}(\alpha_i,\mu) &=& 120 \alpha_u \alpha_s \alpha_g
a_{10} (\alpha_s-\alpha_u)\, ,
\nonumber\\
V_{\perp}(\alpha_i,\mu) &=& -30\alpha_g^2
\left\{h_{00}(1-\alpha_g)+h_{01}\left[\alpha_g(1-\alpha_g)-6\alpha_u
\alpha_s\right] \right.  \nonumber\\
&&\left. +h_{10}\left[
\alpha_g(1-\alpha_g)-\frac{3}{2}\left(\alpha_u^2+\alpha_s^2\right)\right]\right\}\,
, \nonumber\\
A_{\perp}(\alpha_i,\mu) &=&  30 \alpha_g^2 (\alpha_u-\alpha_s) \left\{h_{00}+h_{01}\alpha_g+\frac{1}{2}h_{10}(5\alpha_g-3)  \right\}, \nonumber\\
A(u,\mu)&=&6u(1-u)\left\{
\frac{16}{15}+\frac{24}{35}a_2+20\eta_3+\frac{20}{9}\eta_4 \right.
\nonumber \\
&&+\left[
-\frac{1}{15}+\frac{1}{16}-\frac{7}{27}\eta_3\omega_3-\frac{10}{27}\eta_4\right]C^{\frac{3}{2}}_2(2u-1)
\nonumber\\
&&\left.+\left[
-\frac{11}{210}a_2-\frac{4}{135}\eta_3\omega_3\right]C^{\frac{3}{2}}_4(2u-1)\right\}+\left\{
 -\frac{18}{5}a_2+21\eta_4\omega_4\right\} \nonumber\\
 && \left\{2u^3(10-15u+6u^2) \log u+2\bar{u}^3(10-15\bar{u}+6\bar{u}^2) \log \bar{u}
 \right. \nonumber\\
 &&\left. +u\bar{u}(2+13u\bar{u})\right\} \, ,\nonumber\\
 g_K(u,\mu)&=&1+g_2C^{\frac{1}{2}}_2(2u-1)+g_4C^{\frac{1}{2}}_4(2u-1)\, ,\nonumber\\
 B(u,\mu)&=&g_K(u,\mu)-\phi_K(u,\mu)\, ,
\end{eqnarray}
where
\begin{eqnarray}
h_{00}&=&v_{00}=-\frac{\eta_4}{3} \, ,\nonumber\\
a_{10}&=&\frac{21}{8}\eta_4 \omega_4-\frac{9}{20}a_2 \, ,\nonumber\\
v_{10}&=&\frac{21}{8}\eta_4 \omega_4 \, ,\nonumber\\
h_{01}&=&\frac{7}{4}\eta_4\omega_4-\frac{3}{20}a_2 \, ,\nonumber\\
h_{10}&=&\frac{7}{2}\eta_4\omega_4+\frac{3}{20}a_2 \, ,\nonumber\\
g_2&=&1+\frac{18}{7}a_2+60\eta_3+\frac{20}{3}\eta_4 \, ,\nonumber\\
g_4&=&-\frac{9}{28}a_2-6\eta_3\omega_3 \, ,
\end{eqnarray}
 here  $ C_2^{\frac{1}{2}}(\xi)$, $ C_4^{\frac{1}{2}}(\xi)$,
  $ C_1^{\frac{3}{2}}(\xi)$, $ C_2^{\frac{3}{2}}(\xi)$ and $ C_4^{\frac{3}{2}}(\xi)$ are Gegenbauer polynomials,
  $\eta_3=\frac{f_{3K}}{f_K}\frac{m_q+m_s}{m_K^2}$ and  $\rho^2={(m_s+m_q)^2\over m_K^2}$
 \cite{LCSR,LCSRreview,Belyaev94}.

The explicit expressions of the correlation functions at the level
of quark-gluon degrees of freedom:
\begin{eqnarray}
\Pi_p^{\pi}&=&  \frac{f_\pi m_\pi^2}{m_u+m_d}\int_0^1du
\frac{u\phi_p(u)}{m_d^2-(q+up)^2}-m_df_\pi m_\pi^2\int_0^1du
\int_0^u
dt\frac{u B(t)}{\left\{m_d^2-(q+up)^2\right\}^2} \nonumber\\
&&-\frac{1}{6}\frac{f_\pi m_\pi^2}{m_u+m_d}\int_0^1du
\phi_\sigma(u)\left\{\left[1-u\frac{d}{du}
\right]\frac{1}{m_d^2-(q+up)^2}+\frac{2m_d^2}{\left[m_d^2-(q+up)^2\right]^2}\right\}
  \nonumber\\
  &&+m_df_\pi \int_0^1 du \left\{ \frac{\phi_\pi(u)}{m_d^2-(q+up)^2} -\frac{m_\pi^2m_d^2}{2} \frac{A(u)}{\left[m_d^2-(q+up)^2\right]^3}\right\}\nonumber\\
&&+f_{3\pi}\int_0^1dv \int_0^1d\alpha_g \int_0^{1-\alpha_g}d\alpha_d
\varphi_{3\pi}(\alpha_u,\alpha_g,\alpha_d)\nonumber\\
&&\left\{\frac{(2v-3)u m_\pi^2
}{\left[m_d^2-(q+up)^2\right]^2}+2v\frac{d}{du}\frac{1}{m_d^2-(q+up)^2}\right\}\mid_{u=(1-v)\alpha_g+\alpha_d}
\nonumber\\
&&-4m_df_\pi m_\pi^4\int_0^1dv v \int_0^1 d\alpha_g\int_0^{\alpha_g}
d\beta\int_0^{1-\beta}d\alpha
\frac{u\Phi(1-\alpha-\beta,\beta,\alpha)}{\left\{m_d^2-(q+up)^2\right\}^3}\mid_{1-v\alpha_g}
\nonumber \\
&& +4m_df_\pi m_\pi^4\int_0^1 dv\int_0^1
d\alpha_g\int_0^{1-\alpha_g} d\alpha_d
 \int_0^{\alpha_d}d\alpha
\frac{u\Phi(1-\alpha-\alpha_g,\alpha_g,\alpha)}
{\left\{m_d^2-(q+up)^2\right\}^3}\mid_{u=(1-v)\alpha_g+\alpha_d}\nonumber\\
&&+m_d f_\pi m_\pi^2  \int_0^1dv \int_0^1d\alpha_g
\int_0^{1-\alpha_g}d\alpha_d
\frac{\Psi(\alpha_u,\alpha_g,\alpha_d)}{\left\{m_d^2-(q+up)^2\right\}^2}\mid_{u=(1-v)\alpha_g+\alpha_d}
\, ,
\end{eqnarray}

\begin{eqnarray}
\Pi_p^K&=&  \frac{f_K m_K^2}{m_u+m_s}\int_0^1du
\frac{u\phi_p(u)}{m_s^2-(q+up)^2}-m_s f_K m_K^2\int_0^1du \int_0^u
dt\frac{u B(t)}{\left\{m_s^2-(q+up)^2\right\}^2} \nonumber\\
&&-\frac{1}{6}\frac{f_K m_K^2}{m_u+m_s}\int_0^1du
\phi_\sigma(u)\left\{\left[1-u\frac{d}{du}
\right]\frac{1}{m_s^2-(q+up)^2}+\frac{2m_s^2}{\left[m_s^2-(q+up)^2\right]^2}\right\}
  \nonumber\\
  &&+m_sf_K \int_0^1 du \left\{ \frac{\phi_K(u)}{m_s^2-(q+up)^2} -\frac{m_K^2m_s^2}{2} \frac{A(u)}{\left[m_s^2-(q+up)^2\right]^3}\right\}\nonumber\\
&&+f_{3K}\int_0^1dv \int_0^1d\alpha_g \int_0^{1-\alpha_g}d\alpha_s
\varphi_{3K}(\alpha_u,\alpha_g,\alpha_s)\nonumber\\
&&\left\{\frac{(2v-3)u m_K^2
}{\left[m_s^2-(q+up)^2\right]^2}+2v\frac{d}{du}\frac{1}{m_s^2-(q+up)^2}\right\}\mid_{u=(1-v)\alpha_g+\alpha_s}
\nonumber\\
&&-4m_sf_K m_K^4\int_0^1dv v \int_0^1 d\alpha_g\int_0^{\alpha_g}
d\beta\int_0^{1-\beta}d\alpha
\frac{u\Phi(1-\alpha-\beta,\beta,\alpha)}{\left\{m_s^2-(q+up)^2\right\}^3}\mid_{1-v\alpha_g}
\nonumber \\
&& +4m_sf_K m_K^4\int_0^1 dv\int_0^1 d\alpha_g\int_0^{1-\alpha_g}
d\alpha_s
 \int_0^{\alpha_s}d\alpha
\frac{u\Phi(1-\alpha-\alpha_g,\alpha_g,\alpha)}
{\left\{m_s^2-(q+up)^2\right\}^3}\mid_{u=(1-v)\alpha_g+\alpha_s}\nonumber\\
&&+m_s f_K m_K^2  \int_0^1dv \int_0^1d\alpha_g
\int_0^{1-\alpha_g}d\alpha_s
\frac{\Psi(\alpha_u,\alpha_g,\alpha_s)}{\left\{m_s^2-(q+up)^2\right\}^2}\mid_{u=(1-v)\alpha_g+\alpha_s}
\, ,
\end{eqnarray}

\begin{eqnarray}
\Pi_q^{\pi}&=&  \frac{f_\pi m_\pi^2}{m_u+m_d}\int_0^1du
\frac{\phi_p(u)}{m_d^2-(q+up)^2}-m_df_\pi m_\pi^2\int_0^1du \int_0^u
dt\frac{B(t)}{\left\{m_d^2-(q+up)^2\right\}^2} \nonumber\\
&&+\frac{1}{6}\frac{f_\pi m_\pi^2}{m_u+m_d}\int_0^1du
\phi_\sigma(u)\frac{d}{du} \frac{1}{m_d^2-(q+up)^2}
  \nonumber\\
  &&+f_{3\pi}m_\pi^2\int_0^1dv \int_0^1d\alpha_g
  \int_0^{1-\alpha_g}d\alpha_d
\varphi_{3\pi}(\alpha_u,\alpha_g,\alpha_d)\frac{2v-3
}{\left\{m_d^2-(q+up)^2\right\}^2}\mid_{u=(1-v)\alpha_g+\alpha_d}
\nonumber\\
&&-4m_df_\pi m_\pi^4\int_0^1dv v \int_0^1 d\alpha_g\int_0^{\alpha_g}
d\beta\int_0^{1-\beta}d\alpha
\frac{\Phi(1-\alpha-\beta,\beta,\alpha)}{\left\{m_d^2-(q+up)^2\right\}^3}\mid_{1-v\alpha_g}
\nonumber \\
&& +4m_df_\pi m_\pi^4\int_0^1 dv\int_0^1
d\alpha_g\int_0^{1-\alpha_g} d\alpha_d
 \int_0^{\alpha_d}d\alpha
\frac{\Phi(1-\alpha-\alpha_g,\alpha_g,\alpha)}{\left\{m_d^2-(q+up)^2\right\}^3}\mid_{u=(1-v)\alpha_g+\alpha_d}
 \, ,\nonumber \\
\end{eqnarray}

\begin{eqnarray}
\Pi_q^K&=&  \frac{f_K m_K^2}{m_u+m_s}\int_0^1du
\frac{\phi_p(u)}{m_s^2-(q+up)^2}-m_sf_K m_K^2\int_0^1du \int_0^u
dt\frac{B(t)}{\left\{m_s^2-(q+up)^2\right\}^2} \nonumber\\
&&+\frac{1}{6}\frac{f_K m_K^2}{m_u+m_s}\int_0^1du
\phi_\sigma(u)\frac{d}{du} \frac{1}{m_s^2-(q+up)^2}
  \nonumber\\
  &&+f_{3K}m_K^2\int_0^1dv \int_0^1d\alpha_g
  \int_0^{1-\alpha_g}d\alpha_s
\varphi_{3K}(\alpha_u,\alpha_g,\alpha_s)\frac{2v-3
}{\left\{m_s^2-(q+up)^2\right\}^2}\mid_{u=(1-v)\alpha_g+\alpha_s}
\nonumber\\
&&-4m_sf_K m_K^4\int_0^1dv v \int_0^1 d\alpha_g\int_0^{\alpha_g}
d\beta\int_0^{1-\beta}d\alpha
\frac{\Phi(1-\alpha-\beta,\beta,\alpha)}{\left\{m_s^2-(q+up)^2\right\}^3}\mid_{1-v\alpha_g}
\nonumber \\
&& +4m_sf_K m_K^4\int_0^1 dv\int_0^1 d\alpha_g\int_0^{1-\alpha_g}
d\alpha_s
 \int_0^{\alpha_s}d\alpha
\frac{\Phi(1-\alpha-\alpha_g,\alpha_g,\alpha)}{\left\{m_s^2-(q+up)^2\right\}^3}\mid_{u=(1-v)\alpha_g+\alpha_s}
\, , \nonumber \\
\end{eqnarray}
where $\Psi=A_\parallel-V_\parallel-2A_\perp+2V_\perp$ and
$\Phi=A_\perp+A_\parallel-V_\perp -V_\parallel$.

\section*{Acknowledgments}
This  work is supported by National Natural Science Foundation,
Grant Number 10405009,  and Key Program Foundation of NCEPU.

\end{document}